\documentstyle[prl,multicol,aps]{revtex}
\begin{document}

\draft
\title{A natural orbital functional for the many-electron problem}
\author{S. Goedecker}
\address{ Max-Planck Institute for Solid State Research, Stuttgart, Germany}
\author{C. J. Umrigar}
\address{Cornell Theory Center and Laboratory of Atomic and Solid State,
 Cornell University, Ithaca, NY 14853}
\date{\today}
\maketitle

\begin{abstract}
The exchange-correlation energy in Kohn-Sham density functional theory
is expressed as a functional of the electronic
density and the Kohn-Sham orbitals.
An alternative to Kohn-Sham theory is to express the energy as a functional
of the reduced first-order density matrix or equivalently the natural orbitals.
In the former approach the unknown part of the functional contains both
a kinetic and a potential contribution whereas in the latter approach
it contains only a potential energy and consequently has simpler scaling properties.
We present an approximate, simple and parameter-free functional of the natural orbitals,
based solely on scaling arguments and the near satisfaction of a sum rule.
Our tests on atoms show that it yields
on average more accurate energies and charge densities than the Hartree Fock method,
the local density approximation and the generalized gradient approximations.

\end{abstract}

\begin{multicols}{2}
\setcounter{collectmore}{5}
\raggedcolumns

The solution of the quantum mechanical many-electron problem is one of the central
problems of physics. A great number of schemes that approximate the
intractable many-electron Schr\"{o}dinger equation have been devised to
attack this problem. Most of them map the many-body problem to a self-consistent
one-particle problem. Probably the most popular method at present is
Density Functional Theory~\cite{dft} (DFT) especially when employed with the
Generalized Gradient Approximation~\cite{PBE,BLYP}
(GGA) for the exchange-correlation energy.
DFT is based on the Hohenberg-Kohn theorem~\cite{HK} which asserts that the
electronic charge density completely determines a many-electron system and that
in particular the total energy is a functional of the charge density.
Attempts to construct such a functional for the total energy have not been very
successful because of the strong non-locality of the kinetic energy term.
The Kohn-Sham scheme~\cite{KS} where the main part of the kinetic energy, the single
particle kinetic energy, is calculated by solving one-particle Schr\"{o}dinger
equations circumvented this problem. The difference of the
one-particle kinetic energy and the many-body kinetic energy is
a component of the unknown exchange-correlation functional. The exchange-correlation
functional is thus a sum of a kinetic energy contribution and a potential energy
contribution and partly for this reason it does not scale
homogeneously~\cite{LevyPerdew85} under a uniform
spatial scaling of the charge density.

It has been known for a long time, that one can also construct a total energy
functional using the first-order reduced density matrix.
Several discussions of the existence and the properties of such a functional
can be found in the literature~\cite{Levy79,Levy87,DonellyParr78,Zumbach85}.
However in spite of the enthusiasm
expressed towards this approach in the early papers, no
explicit functional has ever been constructed and tested on real physical
systems.
An important advantage of this
approach is that one employs an exact expression for the many-body kinetic energy.
Only the small non Hartree-Fock-like part of the electronic repulsion
is an unknown functional~\cite{Levy87}. We propose in this paper an explicit
form of such a functional in terms of the natural orbitals. The high accuracy of this
Natural Orbital Functional Theory (NOFT) is then established by applying it
to several atoms and ions.

Let us first briefly review some basic facts about
reduced density matrices~\cite{Lowdin55,Davidson76}.
If $\Psi$ is an arbitrary trial wave function of an $N$-electron system, the
first and second order reduced density matrices, $\gamma_1$ and $\gamma_2$ are
\begin{eqnarray}  \label{red1}
\gamma_1({\bf x}_1',{\bf x}_1) &=& N \int ... \int
     \Psi  ({\bf x}_1',{\bf x}_2, ... ,{\bf x}_N) \nonumber \\
& & \Psi  ({\bf x}_1, {\bf x}_2, ... ,{\bf x}_N) \; d{\bf x}_2 ... d{\bf x}_N, \\
\gamma_2({\bf x}_1',{\bf x}_2';{\bf x}_1,{\bf x}_2) &=& {N(N-1) \over 2} \;
 \int ...  \int  \Psi  ({\bf x}_1',{\bf x}_2',{\bf x}_3, ... ,{\bf x}_N)   \nonumber \\
& & \Psi  ({\bf x}_1,{\bf x}_2,,{\bf x}_3, ... ,{\bf x}_N) \; d{\bf x}_3 ... d{\bf x}_N .
\end{eqnarray}

The variables ${\bf x}_i$ contain both
the position coordinates ${\bf r}_i$, as
well as the spin coordinate $s_i$. The
integration sign stands for a combined integration
of the spatial coordinates and summation of the discrete spin part.

The electronic charge density $\rho({\bf r})$ is obtained from the diagonal
part of the first-order reduced density matrix,
\begin{equation} \label{rho}
\rho({\bf x}_1) = \gamma_1({\bf x}_1,{\bf x}_1) ; \;\;\; \rho({\bf r}_1) = \sum_{s_1} \rho({\bf x}_1).
\end{equation}

The natural orbitals $\phi_i$ are the eigenfunctions of the first-order reduced
density matrix with eigenvalues $n_i$.
\begin{equation}  \label{nos}
 \int \gamma_1({\bf x}_1',{\bf x}_1)  \phi_i({\bf x}_1) d{\bf x}_1 = n_i \phi_i({\bf x}_1')
\end{equation}
The natural spin-orbitals and occupation numbers $n_i$ specify the reduced first-order
density matrix completely.

The total energy can be written in terms of the natural orbitals and
the diagonal elements of the second order reduced density matrix,
\begin{equation} \label{sigma}
\sigma({\bf x}_1,{\bf x}_2) = \gamma_2({\bf x}_1,{\bf x}_2;{\bf x}_1,{\bf x}_2),
\end{equation}
as
\begin{eqnarray}  \label{etot}
 E & = & -\frac{1}{2} \sum_i n_i \int \phi_i({\bf x}) \nabla^2 \phi_i({\bf x}) d{\bf x}  \\
   & + & \int V({\bf x}) \rho({\bf x}) d{\bf x}
    + \int  \int  \frac{ \sigma({\bf x}_1,{\bf x}_2) }{|{\bf r}_1-{\bf r}_2|} d{\bf x}_1 d{\bf x}_2 \; . \nonumber
\end{eqnarray}

In order to construct a natural orbital functional,
it remains to find an approximation for $\sigma$
in terms of the natural orbitals and occupation numbers.
In the following, we assume the standard case of a Hamiltonian that is not spin dependent.
Each natural orbital can then be chosen to be either purely spin up or spin down and can be
labeled by an orbital index $i$ and a spin index $s_i$.

The approximate $\sigma$ we propose has the following form:
\begin{eqnarray}  \label{approx}
    \sigma  [\{n\},\{\phi\}]  =
 &   \sum'_{i,j} & \frac{n_i n_j}{2}
   \phi_i^2({\bf r}_1) \phi_j^2({\bf r}_2) \\
- & \sum'_{i,j} & \frac{\sqrt{n_i n_j}}{2} \delta_{s_i,s_j}
    \phi_i({\bf r}_1) \phi_j({\bf r}_1) \phi_i({\bf r}_2) \phi_j({\bf r}_2) \; . \nonumber
\end{eqnarray}
The primes indicate that the $i=j$ terms are omitted.
To find the ground state, we minimize the functional with respect to both the
natural orbitals and the occupation numbers, under the constraint that the natural
orbitals be orthogonal~\cite{Goedecksic97}.
The functional derivatives are
\begin{eqnarray}  \label{lagpsi}
\frac{\partial E}{\partial \phi_i({\bf r})} & = &
  - \frac{n_i}{2} \nabla^2 \phi_i({\bf r}) +  n_i V({\bf r}) \phi_i({\bf r})  \\
 & + & \sum'_{j}  n_i n_j \phi_i({\bf r}) \int  \frac{ \phi_j^2({\bf r}') }{|{\bf r}-{\bf r}'|}
 \; d {\bf r}' \nonumber \\
 & - & \sum'_{j} \sqrt{n_i n_j} \delta_{s_i,s_j} \phi_j({\bf r}) \int  \frac{ \phi_i(
{\bf r}') \phi_j({\bf r}') }{|{\bf r}-{\bf r}'|} \; d{\bf r}' \; , \nonumber
\end{eqnarray}

\begin{eqnarray}  \label{lagn}
\frac{\partial E}{\partial n_i}  & = &
  -\frac{1}{2} \int \phi_i({\bf r}) \nabla^2 \phi_i({\bf r}) d{\bf r} +
  \int V({\bf r}) \; \phi_i^2({\bf r}) \; d{\bf r} \\
 & + & \sum'_{j}   n_j \int \int
  \frac{ \phi_j^2({\bf r}') \phi_i^2({\bf r}) } {|{\bf r}-{\bf r}'|} \; d{\bf r} d{\bf r}' \nonumber
 \\
 & - & \frac{1}{2} \sum'_{j} \sqrt{\frac{n_j}{n_i}} \delta_{s_i,s_j} \int  \int
   \frac{ \phi_i({\bf r}') \phi_j({\bf r}') \phi_i({\bf r}) \phi_j({\bf r}) }{|{\bf r}-{\bf r}'|}
 d{\bf r} d{\bf r}' \; . \nonumber
\end{eqnarray}

In principle an infinite number of natural orbitals must
be included. For the systems studied in Table~\ref{results} at most 38 orbitals were
needed to obtain good convergence.
The occupation numbers of the core natural orbitals are
restricted to be unity, while the remaining occupation numbers are allowed to vary freely
and are found to lie always between zero and one, which is a necessary
and sufficient condition for the density matrix to be N-representable~\cite{Davidson76}.

We now discuss the properties of this functional.

Homogeneous scaling of exchange-correlation energy: \newline
       The exact exchange-correlation energy in first-order density matrix functional
       theory differs from the exact exchange-correlation energy in density functional
       theory amd scales homogeneously~\cite{Levy87} under a uniform
       scaling of the density matrix.  The exchange-correlation energy, deduced from
       Eqs.~(\ref{etot}) and (\ref{approx}), exhibits this property.

No orbital self-interactions: \newline
       In the case where one has fractional occupation numbers one has to distinguish
       between orbital self-interactions and electron self-interactions. Our functional
       is free of orbital self-interactions because the sum in Eq.~\ref{approx} excludes
       terms with $i=j$, but it is not perfectly electron
       self-interaction free. The total energy for H is therefore not
       correct (Table~\ref{results}). The functional has however a much better cancellation
       of electron self-interactions than density functionals, as can
       be seen from the fact that negative ions are stable (Table~\ref{results}).
       In contrast LDA and GGA bind only a fraction of an additional electron.

Sum rule for second order reduced density: \newline
      The density and the number of electron pairs are obtained by integrating the
      exact second order reduced density matrix.
      \begin{eqnarray}  \label{g2norm}
       \int  \sigma({\bf r}_1,{\bf r}_2) \; d{\bf r}_2 =  \frac{(N-1)}{2} \; \rho({\bf r}_1) \; , \label{single} \\
       \int  \int   \sigma({\bf r}_1,{\bf r}_2) \; d{\bf r}_1 d{\bf r}_2 = \frac{N(N-1)}{2} \; . \label{pair}
      \end{eqnarray}
      Our approximation for the second order reduced density matrix would satisfy
      these equations if the sums in Eq.~\ref{approx} also included the $i=j$ terms.
      We omit these terms because we find that an exact cancellation of the orbital self-interactions
      is more important than an exact fulfillment of the sum rules in Eqs.~(\ref{single}) and (\ref{pair}).
      The sum rules are
      violated only by terms of the order of $n_i (1-n_i)$, which for most systems are small
      since all the occupation numbers are close to either zero or one.

Hartree Fock as limiting case: \newline
      The functional coincides with the Hartree Fock (HF) functional if one
      imposes the additional constraint, that the occupations numbers
      all be 1 or 0.

No dissociation problems: \newline
     Even though the functional contains terms which are similar to the
     HF functional, it should not suffer from some well
     established deficiencies of the spin restricted HF functional such as the
     dissociation problem of the H$_2$ molecule. As one separates the two H atoms,
     the large occupation numbers in the up- and down-spin
     $\sigma_g$ molecular orbital
     get redistributed to the up-spin 1s atomic orbital on one atom and the
     down-spin 1s atomic orbital on the other.  In the infinitely separated limit
     each atom has non-zero occupation numbers in either only the up-spin or only the
     down-spin orbitals.
     Consequently the energy is the sum of the energies of the individual atoms.

Transition states: \newline
      In molecular calculations the effect of this functional is
      expected to be particularly significant for transitions states, which
      are poorly described by LDA and HF.
      At transition states more than one determinant is needed for an adequate
      description, and releasing the HF constraint of integer occupation numbers
      is therefore important.

Orbital-dependent ``potentials": \newline
       The weakly-occupied natural orbitals are localized in the same region of space
       as the highest strongly-occupied natural orbitals.  This is in contrast to
       the unoccupied Kohn-Sham and Hartree-Fock orbitals which have a larger extent
       than the occupied ones.
       The manner in which this comes about can be seen from
       Eq. \ref{lagpsi} which has an orbital-dependent ``potential".
       One term in the potential goes as $\sqrt{n_i}$ -- an enhancement by a factor
       of $1/\sqrt{n_i}$ relative to Hartree-Fock -- which has the consequence that
       weakly-occupied natural orbitals see a more strongly negative potential than
       do the strongly-occupied orbitals, thereby helping to localize the
       weakly-occupied natural orbitals.

Chemical potential: \newline
      All natural orbitals with fractional
      occupation $n_i$ share the same
      chemical potential~\cite{DonellyParr78},
      $ \mu = \frac{\partial E}{\partial n_i} \: .$

Discontinuity of the exchange-correlation potential: \newline
       As one adds fractions of an electron, one finds, at occupation numbers
       close to integers, a rapid change in the
       effective potential felt by all the electrons, which is due to the jump in the
       chemical potential. This quasi discontinuous effect might mimic the
       discontinuity~\cite{PerdewLevy83,JonesGunnarson89} in the DFT exchange
       correlation potential,
       an effect missing in the LDA and GGA functionals.

Correct description of correlations of different origin: \newline
        In a $1/Z$ expansion of the energy, the correlation energy of the two-electron
        series can be described by nondegenerate perturbation theory
        while the four-electron series requires degenerate perturbation theory.
        Consequently the correlation energy of the two-electron series tends to a constant
        with increasing Z, whereas it increases linearly in the four-electron case.
        Both trends are correctly captured by the NOFT functional as shown in Table~\ref{cor}.
        Any GGA functional can at best describe only one of the trends.

Correct qualitative behavior of natural occupation numbers: \newline
        As seen from Table~\ref{nocc}, the NOFT occupation numbers may differ
        considerably from the ones
        obtained from configuration interaction calculations, but
        the main trends are correctly reproduced. In particular, the trend in the
        occupation numbers of the strongly occupied 1s orbitals, going from He to
        H$^-$ is correct.

Accurate results: \newline
       In Table~\ref{results}, we give a compilation
       of the errors in the total energy $\Delta E$ and the errors in the charge densities
       $\Delta \rho$.  The charge density errors are defined by
       $ (\Delta \rho)^2 = \int (\rho_{\rm ex}({\bf r})-\rho({\bf r}))^2 d{\bf r}$,
       with the ``exact" charge densities $\rho_{\rm ex}$ obtained from accurate
       quantum Monte Carlo calculations~\cite{Cyrus}.
       Both total energies and charge densities are improved on average compared to
       HF and DFT calculations. In particular the improvements over the
       HF densities are impressive since they are known to be
       rather accurate. The GGA schemes yield improved total energies compared
       to both LDA and HF while the GGA densities are better than those
       from LDA but not as good as those from HF.
       In the case of C, the error in the spherically
       averaged charge density is quoted.  The ``exact" total energies were obtained
       from Ref.~\onlinecite{atom}.
       The LDA energies and densities were obtained by a standard spherical atomic program.
       As a representative of a GGA functional we have chosen the recent PBE~\cite{PBE}
       functional.
       All the HF and NOFT calculations were done with
       a non-spherical atomic program developed by the authors. All calculations
       were done in a spin restricted scheme.
       In the case of $C$ the correct non-spherical $^3$P ground state was
       chosen. 
       The QCISD configuration interaction calculations were done with the
       Gaussian 94 software package~\cite{g94} using an accurate 6-311G++G(3df,2p)
       basis set.
       Since we do no molecular calculations, we monitor a third quantity
       the transferability error $\Delta \tau$, to make predictions about the behavior
       of this scheme in molecular and solid state calculations.
       Molecular geometries are
       determined via the Hellmann-Feynman theorem by the charge densities in the
       valence region. The external potential in the valence region is modified in
       a molecule compared to the atomic case. We simulated this modification
       by adding a confining parabolic potential to the atom. The change in the
       total energy due to the variation of this parabolic potential
       is again given by the Hellmann-Feynman theorem and we
       define the transferability error $\Delta \tau$ therefore as
       $ \Delta \tau = \int (\rho({\bf r})-\rho_{\rm ex}({\bf r})) \; r^2 d{\bf r}$.

\bigskip
In conclusion, we have made a first attempt at constructing an approximate
total energy functional of the first-order reduced density matrix.
We have listed and discussed the properties
that make it superior to the HF and approximate DFT functionals and
have also shown that it yields better energies and densities than HF and current DFT schemes.
The high
accuracy of quantities related to the charge density leads one to expect that this
new functional will give accurate molecular geometries as well
as accurate energy differences between different geometric configurations.
In view of the fact that
the functional is parameter free and based on a few simple
considerations, we think this to be a remarkable success. It is likely
that it will be possible to construct even better functionals along these lines.
The essential point in this work is that we have used natural orbitals instead of
Kohn Sham orbitals. We believe that this is essential
to obtain accurate densities and kinetic energies.
With the exception of the complication of an orbital-dependent potential,
the computational procedure necessary
to solve the NOFT equations is analogous to other self-consistent
one-particle schemes and thus computationally much cheaper than quantum chemistry
methods based on configuration interaction related schemes.

We thank M. Levy, O. Gunnarson, J. Hutter, M. Teter, K. Maschke, A. Savin and B. Fahrid for
interesting discussions and for suggesting references.
Mike Harris kindly provided a subroutine for angular grid generation.

\begin{minipage}{3.375in}
\begin{table}
\caption[]{\label{results} Comparison of the errors of the quantities described in the
text.  Energies are in Hartree atomic units.
No data are available (NA) for the non-spherical PBE ground state of C.
The large errors in $\Delta \rho$ and the infinite errors in $\Delta \tau$ for the H$^-$ ion
in LDA and PBE come from the fact that they bind only a fraction of the additional electron.}
\begin{tabular}{|l||c|c|c|c|c|c|c|}
                &   H   &  H$^-$ &  He   &  Li   &  Be   &  C    &   Ne    \\ \hline
                &    \multicolumn{7}{c}{Energy}                            \\ \hline
  - E           &  .5   & .5278  &2.9037 &7.4781 &14.6674&37.8450&128.9376 \\ \hline
                &    \multicolumn{7}{c}{LDA}                               \\ \hline
 $\Delta E$     & 2.e-2 & 6.e-3  & 7.e-2 & 1.e-1 & 2.e-1 & 4.e-1 & 7.e-1   \\ \hline
$(\Delta \rho)^2$&1.e-3 & 6.e0   & 8.e-3 & 2.e-2 & 2.e-2 & 5.e-2 & 2.e-1   \\ \hline
 $\Delta \tau $ & 4.e-1 &$\infty$& 2.e-1 &-7.e-1 & 2.e-2 & 4.e-1 & 3.e-1   \\ \hline
                &    \multicolumn{7}{c}{PBE}                               \\ \hline
 $\Delta E$     & 8.e-5 & 2.e-3  & 1.e-2 & 2.e-2 & 4.e-2 &  NA   & 7.e-2   \\ \hline
$(\Delta \rho)^2$&2.e-4 & 6.e0   & 1.e-3 & 3.e-3 & 3.e-3 &  NA   & 1.e-2   \\ \hline
 $\Delta \tau$  & 2.e-1 &$\infty$& 1.e-1 &-1.e0  & 5.e-1 &  NA   & 3.e-1   \\ \hline
                &    \multicolumn{7}{c}{HF}                                \\ \hline
 $\Delta E$     &   0.  & 4.e-2  & 4.e-2 & 5.e-2 & 9.e-2 & 2.e-1 & 4.e-1   \\ \hline
$(\Delta \rho)^2$&  0.  & 1.e-3  & 1.e-4 & 7.e-5 & 8.e-4 & 5.e-4 & 6.e-3   \\ \hline
 $\Delta \tau$  &   0.  &-5.e0   &-2.e-2 & 3.e-1 & 1.e0  &-6.e-2 &-2.e-1   \\ \hline
                &    \multicolumn{7}{c}{NOFT}                              \\ \hline
 $\Delta E$     &-2.e-2 & 1.e-2  & 6.e-3 &-1.e-3 &-2.e-2 & 3.e-2 & 5.e-2   \\ \hline
$(\Delta \rho)^2$&3.e-5 & 4.e-4  & 1.e-5 & 2.e-4 & 6.e-4 & 7.e-4 & 4.e-4   \\ \hline
 $\Delta \tau$  &-2.e-2 & 1.e2   &-1.e-2 &-5.e-1 & 6.e-1 & 5.e-2 &-5.e-2   \\ \hline
                &    \multicolumn{7}{c}{QCISD}                             \\ \hline
  $\Delta E$    & 2.e-3 & 8.e-3  & 8.e-3 & 5.e-2 & 5.e-2 & 7.e-2 & 1.e-1

\end{tabular}
\end{table}
\end{minipage}

\begin{minipage}{3.375in}
\begin{table}
\caption[]{\label{cor} Correlation energies, in Hartrees, for the 2- and 4-electron series.
The exact values of $-E_c^{\rm QC} = E^{\rm HF} - E^{\rm exact}$, taken from Ref.~\onlinecite{atom},
are compared to  $E^{\rm HF} - E^{\rm NOFT}$.}
\begin{tabular}{|c|c|c||c|c|c|}
       \multicolumn{3}{c}{2 electron}    &  \multicolumn{3}{c}{4 electron}        \\ \hline
Z   & $-E_c^{QC}$ &$E^{\rm HF}-E^{\rm NOFT}$&  Z   & $-E_c^{QC}$ &$E^{\rm HF}-E^{\rm NOFT}$\\ \hline
1   &     .040    &       .031              &  4   &     .094    &    .110                 \\ \hline
2   &     .042    &       .036              &  6   &     .126    &    .141                 \\ \hline
4   &     .044    &       .040              &  8   &     .154    &    .171                 \\ \hline
6   &     .045    &       .042              &  10  &     .180    &    .200
\end{tabular}
\end{table}
\end{minipage}

\begin{minipage}{3.375in}
\begin{table}
\caption[]{\label{nocc} Occupation numbers for the 2-electron series.
 Columns labeled 'E' are the almost exact numbers of Kutzelnigg~\cite{kutzel}.
 Entries smaller than 1e-5 were set to zero. }
\begin{tabular}{|l||c|c|c|c|c|c|c|c|}
 nl     & E:Z=1  &  Z=1   &  E:Z=2  &  Z=2   & E:Z=4  & Z=4    &  E:Z=6 & Z=6     \\ \hline
 1s     & .9646  & .9666  &  .9930  & .9943  & .9984  & .9985  & .9993  & .9993   \\ \hline
 2s     & .24e-1 & .10e-1 &  .32e-2 & .22e-2 & .63e-3 & .53e-3 & .25e-3 & .24e-3  \\ \hline
 2p     & .11e-1 & .28e-2 &  .34e-2 & .92e-3 & .89e-3 & .27e-3 & .39e-3 & .13e-3  \\ \hline
 3s     & .73e-4 & .10e-1 &  .36e-4 & .11e-3 & 0      & .19e-4 & 0      & 0       \\ \hline
 3p     & .15e-3 & .47e-3 &  .79e-4 & .77e-4 & .23e-4 & .16e-4 & 0      & 0       \\ \hline
 4s     & 0      & .27e-3 &  0      & 0      & 0      & 0      & 0      & 0       \\ \hline
 3d     & .37e-3 & .30e-3 &  .15e-3 & .61e-4 & .47e-4 & .13e-4 & .21e-4 & 0       \\ \hline
 4p     & .12e-4 & .11e-3 &  0      & 0      & 0      & 0      & 0      & 0       \\ \hline
 5s     & 0      & .81e-4 &  0      & 0      & 0      & 0      & 0      & 0       \\ \hline
 4d     & .33e-4 & .57e-4 &  .14e-4 & 0      & 0      & 0      & 0      & 0       \\ \hline
 5p     & 0      & .91e-4 &  0      & 0      & 0      & 0      & 0      & 0       \\ \hline
 6s     & 0      & .12e-4 &  0      & 0      & 0      & 0      & 0      & 0       \\ \hline
 6p     & 0      &  0     &  0      & 0      & 0      & 0      & 0      & 0
\end{tabular}
\end{table}
\end{minipage}

\end{multicols}
\end{document}